\documentclass[twocolumn,prl,amssymb,amsmath,preprintnumbers,aps,nofootinbib]{revtex4-1}
\usepackage{lineno}
\usepackage{bm,color,xcolor}
\usepackage{slashed} 
\usepackage{graphicx}
\usepackage{soul} 
\usepackage{multirow}
\usepackage{comment}
\usepackage{epstopdf} 
\usepackage{mathtools}
\usepackage{appendix}
\usepackage{cancel}
\usepackage[colorlinks=true
,urlcolor=blue
,anchorcolor=blue
,citecolor=blue 
,filecolor=blue
,linkcolor=blue
,menucolor=blue
,linktocpage=true
,pdfproducer=medialab
,pdfa=true
]{hyperref}

\newcommand{\beq}{\begin{equation}}
\newcommand{\eeq}{\end{equation}}
\newcommand{\bea}{\begin{eqnarray}}
\newcommand{\eea}{\end{eqnarray}}
\newcommand{\ba}{\begin{array}}
\newcommand{\ea}{\end{array}}

\def\m1{M_1}
\def\m2{M_2}
\def\m3{M_3}

\def\ch10{\tilde \chi^0_1}

\def\to{\rightarrow}

\newcommand{\lsim}{\mathrel{\mathop{\kern 0pt \rlap
  {\raise.2ex\hbox{$<$}}}
  \lower.9ex\hbox{\kern-.190em $\sim$}}}
\newcommand{\gsim}{\mathrel{\mathop{\kern 0pt \rlap
  {\raise.2ex\hbox{$>$}}}
  \lower.9ex\hbox{\kern-.190em $\sim$}}}
\begin{document}
\title{\boldmath \bf \Large 
Hybrid SO(10) Axion Model Without Quality Problem
}

\author{\bf K.S. Babu$^1$, Bhaskar Dutta$^2$, and Rabindra N. Mohapatra$^3$}

\affiliation{$^1$Department of Physics, Oklahoma State University, Stillwater, OK 74078\\$^2$Department of Physics and Astronomy, Mitchell Institute of Fundamental Physics and Astronomy, Texas A$\&$M University, College Station, TX  77843, USA\\$^3$Maryland Center for Fundamental Physics, Department of Physics,
University of Maryland, College Park, MD 20742, USA}
\preprint{MI-HET-840}

\begin{abstract}
Invisible axion models that solve the strong CP problem via the Peccei-Quinn (PQ) mechanism typically have a quality problem that arises from quantum gravity effects which violate all global symmetries. These models therefore require extreme fine-tuning of parameters for consistency. We present a new solution to the quality problem in a unified $SO(10)\times U(1)_a$ gauge model, where $U(1)_a$ is an anomaly free axial gauge symmetry. PQ symmetry emerges as an accidental symmetry in this setup, which admits a PQ breaking scale as large as $4\times 10^{11}$ GeV, allowing for the axion to be the cosmological dark matter. We call this a hybrid axion model due to its unique feature that it interpolates between the popular KSVZ and DFSZ axion models. Its predictions for the experimentally measurable axion couplings to the nucleon and  electron are distinct from those of the usual models, a feature that can be used to test it. Furthermore, the model has no domain wall problem and it provides a realistic and predictive framework for fermion masses and mixings. 

\begin{center}{\large \it This paper is dedicated to the memory of our friend Satyanaryan Nandi}\end{center}
\end{abstract}

\preprint{
}

\maketitle

{\bf Introduction:}  It is well known that the Standard Model (SM) of weak, electromagnetic and strong interactions suffers from an uncontrolled amount of CP violation arising from non-perturbative QCD effects. This is parametrized by the Lagrangian 
\begin{equation}
{\cal L}_{\rm \cancel{CP}}^{\rm QCD}=\frac{g_s^2}{32\pi^2} \,\overline{\theta}\, G_{\mu \nu}^a \tilde{G}^{a,\mu \nu},
\end{equation}
where $G_{\mu \nu}^a$ is the gluon field strength tensor, $\tilde{G}^{a,\mu \nu} = \frac{1}{2}\epsilon^{\mu \nu \alpha \beta} G_{\alpha \beta}^a$ is its dual, and $\overline{\theta}$ is a dimensionless parameter. In presence of ${\cal L}_{\rm \cancel{CP}}^{\rm QCD}$ the neutron would acquire a nonzero electric dipole moment of the order $d_n \sim (10^{-16}\,\overline{\theta})\, e$-cm~\cite{Crewther:1979pi}. The current experimental limit on $d_n$, $|d_n| \leq 1.8 \times 10^{-26}\,e$-cm \cite{Abel:2020pzs}, leads to the stringent constraint $|\overline{\theta}| \leq 10^{-10}$. The unexplained smallness of $|\overline{\theta}|$ is referred to as the strong CP problem, which suggests new  ingredients that go beyond the Standard Model.

The most widely discussed solution to the strong CP problem is based on the Peccei-Quinn (PQ) mechanism~\cite{Peccei:1977hh}, which promotes the $\overline{\theta}$ parameter to a dynamical field. Here one postulates the existence of a light pseudo-scalar boson, the axion $a(x)$, with coupling to the gluon field which modifies ${\cal L}_{\rm \cancel{CP}}^{\rm QCD}$ to 
\begin{equation}
{\cal L}_{\rm \cancel{CP}}^{\rm QCD}=\frac{g_s^2}{32\pi^2} \left(\overline{\theta} + \frac{a(x)}{f_a} \right)G_{\mu \nu}^a \tilde{G}^{a,\mu \nu},
\label{eq:axion}
\end{equation}
where $f_a$ is the axion decay constant, a parameter with mass dimension. The axion field is realized as a pseudo-Nambu-Goldstone boson associated with the spontaneous breaking of a global $U(1)_{\rm PQ}$ symmetry which acts on the quark fields~\cite{Weinberg:1977ma,Wilczek:1977pj}. This $U(1)_{\rm PQ}$ symmetry is explicitly broken by a QCD anomaly which induces the coupling given in Eq. (\ref{eq:axion}). This in turn leads to an axion potential which can be computed within chiral perturbation theory as~\cite{Weinberg:1977ma,DiVecchia:1980yfw}
\begin{equation}
    {\cal V} \approx -m_\pi^2 f_\pi^2 \sqrt{1-\frac{4 m_u m_d} {(m_u+m_d)^2} \sin^2\left(\frac{a(x)}{2 f_a} +\frac{\overline{\theta}}{2} \right)}
    \label{eq:pot}
\end{equation}
where $m_{u,d}$ are the up and down quark masses and $f_\pi$ is the pion decay constant. Minimizing this potential will set $(\overline{\theta} + a/f_a) = 0$, thus solving the strong CP problem dynamically. The axion will also acquire a mass from Eq. (\ref{eq:pot}) given by
\begin{equation}
m_a^2 \approx \frac{m_u m_d}{(m_u+m_d)^2} \frac{m_\pi^2 f_\pi^2}{f_a^2}~.
\end{equation}

The axion field develops couplings to the fermion $f$ through the interaction term
\begin{equation}
{\cal L}_{af} = \frac{\partial_\mu a}{2 f_a}\,\left( C_{af} \overline{f} \gamma^\mu \gamma_5 f\right)
\label{eq:Cf}
\end{equation}
where $C_{af}$ is a flavor- and model-dependent parameter. Models with $f_a$ of order of the electroweak scale are ruled out by laboratory experiments such as $K^+ \rightarrow \pi^+ + a$ decay, while those with higher values are consistent, but $f_a$ is constrained from astrophysical and cosmological considerations to be in the range $f_a = (10^9-10^{12})$ GeV. These viable models with a high scale value of $f_a$ fall into the category of invisible axion models~\cite{Kim:1979if,Shifman:1979if,Dine:1981rt,Zhitnitsky:1980tq}, for recent reviews see Ref.~\cite{DiLuzio:2020wdo, GrillidiCortona:2015jxo}.

Since the axion field arises from the spontaneous breaking of a global $U(1)_{\rm PQ}$ symmetry at a high scale, the invisible axion models come with a price. It is believed that all global symmetries of nature are broken by non-perturbative gravity effects such as black holes and worm holes, which would imply that the $U(1)_{\rm PQ}$-symmetric Lagrangian will also receive explicit breaking terms parameterized by higher dimensional operators suppressed by the Planck scale. These effects would displace $\overline{\theta}$ away from zero significantly and would spoil the strong CP solution, unless the coefficients of 
the relevant operators turn out to be extremely small. For example, the coefficient of a Planck-suppressed dimension-five term violating $U(1)_{\rm PQ}$ symmetry should be $\leq 10^{-50}$ in order to maintain the strong CP solution. 
The severe fine-tuning needed is referred to as the axion quality problem~\cite{Kamionkowski:1992mf,Holman:1992us,Barr:1992qq,Ghigna:1992iv}.

Several solutions to the quality problem have been proposed in the literature.  These employ new mechanisms such as an accidental $U(1)_{\rm PQ}$ symmetry emerging from gauged $U(1)$~\cite{Barr:1992qq,Qiu:2023los} or non-Abelian gauge symmetries~\cite{DiLuzio:2020qio,Ardu:2020qmo}, 
composite  axion ~\cite{Randall:1992ut,Lillard:2018fdt,Vecchi:2021shj,Lee:2018yak,Cox:2023dou, Cox:2021lii,Nakai:2021nyf}, discrete gauge symmetries~\cite{Babu:2002ic}, mirror universe models~\cite{Berezhiani:2000gh, Hook:2019qoh}, multiple replication of the SM~\cite{Hook:2018jle, Banerjee:2022wzk},  and extra dimensional~\cite{Choi:2003wr} and string theoretic constructions~\cite{Svrcek:2006yi}.

We propose in this Letter a solution to the axion quality problem based on a simple $SO(10)$ grand unified theory extended by a $U(1)_a$ gauge symmetry (the subscript $a$ here stands for axial) which leads to an accidental $U(1)_{\rm PQ}$ symmetry.
The gauge structure of the model is such that it restricts the Planck-suppressed operators that violate $U(1)_{\rm PQ}$ to sufficiently high orders so that there is no axion quality problem. Our model is a new high quality grand unified axion model, that is realistic and whose implications for fermion masses and mixings, including neutrino oscillations, have been extensively studied in the literature~\cite{Babu:1992ia,Bajc:2001fe,Fukuyama:2002ch,Bajc:2002iw,Goh:2003sy,Goh:2003hf,Babu:2005ia,Bertolini:2004eq,Bertolini:2005qb,Bertolini:2006pe,Bajc:2008dc,Joshipura:2011nn,Dueck:2013gca,Altarelli:2013aqa, Fukuyama:2015kra,Babu:2018tfi,Ohlsson:2019sja,Babu:2020tnf}. A unique property of this model is that it is a hybrid axion model that interpolates between the popular KSVZ~\cite{Kim:1979if,Shifman:1979if} and DFSZ~\cite{Dine:1981rt,Zhitnitsky:1980tq} axion models and has characteristic experimental predictions of its own for the axion-electron and axion-nucleon couplings, that can be used to test it. 
The quality constraint on the PQ scale allows for the axion to account for the full dark matter content of the universe.

{\bf The model:} Our model is based on the gauge group $SO(10)\times U(1)_a$ with fermions and scalars assigned to its representations as shown in Table I. The second and third columns of the Table list the gauge quantum numbers while the last column lists the charges under an accidental global $U(1)$ symmetry present in the model. The global $U(1)$ is not uniquely determined, as any linear combination of the global $U(1)$ listed in Table I and the gauge $U(1)_a$ is also a good global symmetry of the model. The global $U(1)$ has a QCD anomaly with the anomaly coefficient given by $A[SU(3)_c^2 \times U(1)_{\rm global}] = 6$, and thus can be identified as the PQ symmetry.  
\begin{table}[htbp]
\begin{center}
\begin{tabular}{|c||c||c||c|}\hline 
\textbf{Fermion} & \textbf{\boldmath{$SO(10)$}} & \textbf{\textbf{gauge} \boldmath{$U(1)_a$}} & \textbf{global} \boldmath{ $U(1)$}\\
& \textbf{irrep} & \textbf{charge} &\textbf{charge}\\\hline
$\psi_a$ & {\bf 16}$_a$ & +1 & +1\\
$F$ & {\bf 10} & $-6$ & ~~0\\
$\chi$ & {\bf 1} & $+12$ & ~~0\\
$N_{1,2,3}$ & {\bf 1} &$(-4, -4, +8)$& $$(0,\,0,+2)$$  \\\hline\hline
\textbf{Scalar} & \textbf{\boldmath{$SO(10)$} rep} & \textbf{\boldmath{$U(1)_a$} charge} &  \textbf{global} \boldmath{$U(1)$} \\\hline
$H$ & {\bf 10} & $-2$ & $-2$ \\
$H'$ & {\bf 10} & ~~$0$ & ~~0\\
$\overline{\Delta}$ & $\overline{{\bf 126}}$& $-2$ & ~$-2$\\
$T$ & {\bf 1} & +1& ~+1\\
$S$ & {\bf 1} & +12 & ~~~0 \\
$A$ & {\bf 45/210} & ~0 & ~~~0\\\hline\hline
\end{tabular}
\end{center}
\caption{Fermion and scalar multiplets of the $SO(10) \times U(1)_a$ model. 
There are three copies of ${\bf 16}_a$ fermions, corresponding to three generations. 
The last column lists the accidental global $U(1)$ symmetry present in the model, with a QCD anomaly which will be identified as $U(1)_{\rm PQ}$. 
}
\end{table}

The model with the particle content listed in Table I is the simplest $SO(10)$ model one can write down with an anomaly free and family-universal $U(1)$ gauge symmetry. The {\bf 10}-fermion, $F$, with $U(1)_a$ charge of $-6$ is used to cancel the $SO(10)^2 \times U(1)_a$ anomaly. The singlet fermions $\chi$ and $N_a$ help to cancel the $[U(1)_a]^3$ and the $({\rm gravity})^2 \times U(1)_a$ anomalies.  In the Higgs sector, the ($H,\overline{\Delta}, A$) fields are the usual ones employed in minimal $SO(10)$ models for consistent symmetry breaking and fermion mass generation. To this set, a real ${\bf 10}$-plet Higgs $H'$ is added to avoid a weak scale axion, along with the singlet scalar $T$.  And the singlet scalar $S$ is needed to generate mass for the ${\bf 10}$-plet fermion $F$. It is highly nontrivial that such a setup has an automatic $U(1)_{\rm PQ}$ symmetry that can solve the axion quality problem. 

The Yukawa Lagrangian of the model is given by:
\begin{eqnarray}\nonumber
{\cal L}_{\rm Yuk} &=&\psi^T\left(Y_{10}\,H+Y_{126}\overline{\Delta}\,\right)\psi\
+ FFS + \chi\chi (S^*)^2/M_{\rm Pl}\\\nonumber &+&  FN_3 H+
 \chi N_{1,2}(T^{4}S^*/M^4_{\rm Pl}+T^{*8}/M^7_{\rm Pl})\nonumber\\&+&\sum_{a,b=1,2}N_aN_b(T^{*4}S/M^4_{\rm Pl}+ T^8/M^7_{\rm Pl}) \nonumber \\ &+&N_{1,2}N_3T^{*4}/M^3_{\rm Pl} +h.c.
\label{eq:Yuk}
\end{eqnarray}
The first two terms of Eq. (\ref{eq:Yuk}) with symmetric Yukawa matrices $Y_{10}$ and $Y_{126}$ lead to realistic and predictive neutrino spectrum which is compatible with current observations~\cite{Babu:1992ia,Bajc:2001fe,Fukuyama:2002ch,Bajc:2002iw,Goh:2003sy,Goh:2003hf,Babu:2005ia,Bertolini:2004eq,Bertolini:2005qb,Bertolini:2006pe,Bajc:2008dc,Joshipura:2011nn,Dueck:2013gca,Altarelli:2013aqa, Fukuyama:2015kra,Babu:2018tfi,Ohlsson:2019sja,Babu:2020tnf}.
The third term in Eq. (\ref{eq:Yuk}) induces masses for the {\bf 10}-plet fermions, and the fourth term induces a TeV-scale mass for the singlet fermion $\chi$. The $F N_3 H$ coupling facilitates the decay of the color-triplet fermion in $F$ through exchange of color-triplet scalar from $H$, $F \rightarrow \overline{N}_3 + \overline{t} + \overline{b}$, with a lifetime of order $10^{-11}$ sec., which is compatible with big bang nucleosynthesis (BBN) constraints. The other terms in Eq. (\ref{eq:Yuk}) which arise through Planck-suppressed operators generate sub-eV masses for the singlet fermions $N_a$.

The Higgs potential of the model contains nontrivial terms that can be written symbolically as (for $A={\bf 45}$):
\begin{eqnarray}\nonumber
\label{pot}
V &\supset& HH'T^2+
 H \Delta AA+ \Delta \Delta HH+\Delta \bar{\Delta} \Delta H  \nonumber \\&+& \Delta \bar{\Delta} HH^* 
 +\Delta \bar{\Delta}\Delta \bar{\Delta}+T^{12}S^*/M^9_{\rm Pl}+ h.c.
 \label{eq:pot1}
\end{eqnarray}
Here the $\Delta \bar{\Delta} \Delta H$ coupling induces mixing between the Higgs doublets of {\bf 10}
and {\bf 126} needed for realistic fermion spectrum. The term $H H' T^2$ is crucial to avoid a weak-scale axion. The Higgs potential has a global $U(1)_{\rm PQ}$ symmetry with the charges listed in the fourth column of Table 1, which is however broken by the Planck-suppressed operator in the last term of Eq. (\ref{eq:pot1}). Due to its high dimensionality its contribution to the axion mass and to the shift in $\overline{\theta}$ are very small, which is why the model has no axion quality problem.  (The higher dimensional terms involving singlet fermion fields $N_a$ of Eq. (\ref{eq:Yuk}) also break the global $U(1)_{\rm PQ}$, but these breakings have no bearing on the axion quality.)

Symmetry breaking in the model proceeds as follows. The {\bf 45/210} scalar and the $(B-L)=2$ component of the {\bf 126}-plet field jointly break the $SO(10)$ gauge symmetry down to that of the SM, while preserving the $U(1)_a \times U(1)_{\rm global}$ symmetry. The singlet fields $S$ and $T$, which acquire VEVs of order $(10^9-10^{12})$ GeV break these surviving symmetries.  Unification of gauge couplings and the rates for proton decay in this model are similar to those discussed in~Ref. \cite{Bertolini:2009qj,Babu:2016bmy,Babu:2015bna,Jarkovska:2021jvw}. The $(B-L)$ symmetry breaking occurs in the model at an intermediate scale $M_I\simeq 10^{12}$ GeV, which is also the seesaw scale for neutrino masses.

{\bf The axion field:} In order to analyze the properties of the axion in our model we first identify its composition in terms of the imaginary components of the complex scalar fields of the model. Not including the SM singlets from ${\bf 45/210}$ field which have no contribution to the axion, the model has three SM singlets denoted as $(\Delta_R^{\overline{126}},\, S,\,T)$ and five weak doublets, $H_{u,d}$ from ${\bf 10}(H)$, 
$H_{u,d}'$ from $\overline{\bf 126}$ 
and $\tilde{H}$ from ${\bf 10}(H')$. The axion is in general a linear combination of all the imaginary components of these fields. We can express these complex fields in the exponential parametrization as $\phi =  (v_\phi e^{i \alpha_\phi}/\sqrt{2})\,e^{i\frac{\eta_\phi(x)}{v_\phi}}$ where $(v_\phi,\,\alpha_\phi)$ are real parameters and $\eta_\phi(x)$ is a dynamical field. We denote the $v_\phi$ values as $(v_R, \,v_S, \,v_T)$  for the SM singlet fields and as $(v_u,\, v_d,\, v_u',\, v_d',\,\tilde{v}$) for the doublets in an obvious notation (shown explicitly in Eq. (\ref{eq:fields}) of Appendix A). The axion field, which is a linear combination of $(\eta_R, \,\eta_S, \,\eta_T, \,\eta_u,\, \eta_u', \,\eta_d, \eta_d',\, \tilde{\eta})$ should be orthogonal to the three Goldstone bosons eaten up by the massive neutral gauge bosons $(Z_\mu,\,X_\mu,\,V_\mu^{(a)})$, as well as four massive pseudoscalar Higgs fields $A_i$.  The composition of these fields are given in Eqs. (\ref{eq:Gold})-(\ref{eq:pseudo}) of Appendix A, from which we obtain the axion field to be
\begin{eqnarray}
a = N( c_S \eta_S + c_T \eta_T+ c_u \eta_u +c_u' \eta_u' + c_d \eta_d + c_d' \eta_d' + \tilde{c} \tilde{\eta}).~
\label{eq:afield}
\end{eqnarray}
The coefficients appearing in Eq. (\ref{eq:afield}) can be written using the definitions
\begin{eqnarray}
&& V_u^2 =v _u^2 + v_u'^2,~~V_d^2 = v_d^2 + v_d'^2,~~v^2 = V_u^2+V_d^2+\tilde{v}^2 \nonumber \\
&& X = v_T^2 \,v^2 + 4\,\tilde{v}^2(V_u^2+V_d^2) + 16 \,V_u^2 V_d^2 \nonumber \\
&&\tan\beta_u = \frac{v_u'}{v_u}, ~\tan\beta_d = \frac{v_d'}{v_d}
\end{eqnarray}
with $v^2 = (174~{\rm GeV})^2$ as
\begin{eqnarray}
&&c_S = X,~~ c_T = -12\, v_S\, v_T \,v^2 \nonumber \\
&&c_u = 24\, v_S V_u(2\, V_d^2 + \tilde{v}^2) \cos\beta_u \nonumber \\
&&c_u' = 24 \,v_S V_u(2\, V_d^2+ \tilde{v}^2) \sin\beta_u\nonumber \\
&&c_d = 24 \,v_S V_d(2 \,V_u^2 + \tilde{v}^2) \cos\beta_d \nonumber \\
&&c_d' = 24 \,v_S V_d (2\,V_u^2+\tilde{v}^2) \sin\beta_d \nonumber\\
&&\tilde{c} = 24\, v_S \tilde{v} (V_d^2-V_u^2)\nonumber\\
&&N = 1/\sqrt{X(X+ 144 v_S^2 v^2)}~.
\end{eqnarray}
Note that the $\eta_R$ field disappears from $a$. 
Note also that since $v^2 \ll v_S^2, v_T^2$ we have $X \simeq v_T^2 \,v^2$ and $N \simeq 1/(v_T^2 \,v^2 \sqrt{1+ 144 v_S^2/v_T^2})$.

To see that $a$ plays the role of axion, we need to find its coupling to gluons. This comes about because of the $\eta_{u,d}$ and $\eta_{u',d'}$ content of Eq. (\ref{eq:afield}) through their couplings to the quarks in {\bf 16}-fermions as well as from the $\eta_S$ component through its couplings to the quarks in the {\bf 10}-fermion. A straightforward calculation shows
\begin{eqnarray}
{\cal L}_{a G\tilde{G}} = \frac{\alpha_s }{8\pi f_a} a\, G_{\mu \nu}^a\tilde{G}^{a, \mu \nu},~~
f_a= v_S /\sqrt{1+\frac{144 v^2_Sv^2}{X}}.~~~
\label{eq:fa}
\end{eqnarray}
\begin{figure}[t!]
		\centering
		\includegraphics[width=0.49\textwidth]{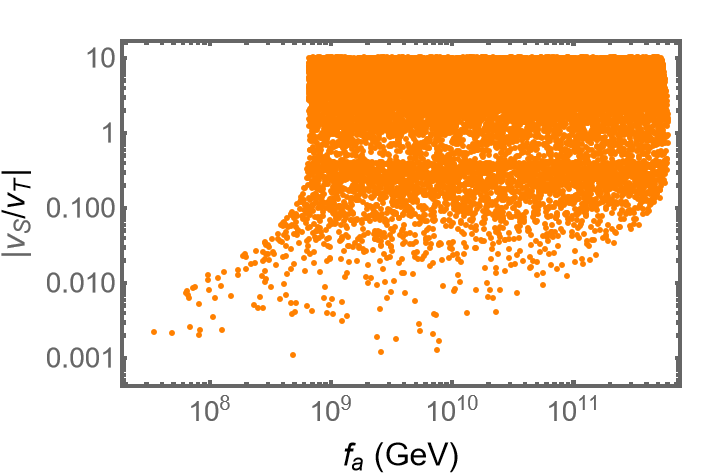}
	\caption{Allowed values of $f_a$ as a function of the VEV ratio $v_S/v_T$.  here we have imposed the axion quality constraint on the parameters of the model.}
	\label{Fig:1}
\end{figure}

{\bf Axion quality constraint:}
From the gauge quantum numbers of the various fields in the model (see Table I), we find that the leading gravity-induced PQ symmetry breaking term in the scalar potential is
\begin{equation}
V_{\rm gravity} = \frac{\kappa\, e^{i \delta}\, T^{12}\, S^*}{(12)!\, M_{\rm Pl}^9} + h.c.
\end{equation}
This shifts the minimum of the axion potential from $\overline{\theta}=0$ to a finite value. 
Defining 
  $r\equiv\frac{v_S^2}{v_T^2}$
and using the relation $X \simeq v_T^2 v^2$, the shift in $\overline{\theta}$ is found to be
\begin{eqnarray}
\overline{\theta} \simeq \left(\frac{\kappa \sin\delta}{(12)!\,2^{11/2}}\right) \left(\frac{f_a^{13}}{f_\pi^2\, m_\pi^2\, M_{\rm Pl}^9}\right)\nonumber\\ \times \frac{(m_u+m_d)^2} {m_u\, m_d} \left(\frac{(1+144 r)^{13/2}}{r^6}\right)~.
\end{eqnarray}
The minimum of $\overline{\theta}$ as a function of $r$ occurs for $r=1/12$.  Using this, and using $\kappa \sin\delta = 1$, $M_{\rm Pl} = 1.22 \times 10^{19}$ GeV and $m_u/m_d = 0.55$, the maximum value for $f_a$ that keeps $|\overline{\theta}| \leq 10^{-10}$ is found to be $f_a < 4.2\times 10^{11}$ GeV. This can also be seen from Fig. 1, which shows a scatter plot of  $f_a$  obtained by varying the weak scale VEVs $V_{u,d}$, the angles ($\beta_u,\,\beta_d$)  and the high scale VEVs $v_{S,T}$. This upper limit on $f_a$ corresponds to a lower bound on the axion mass, $m_a\geq 14$ $\mu$eV. In obtaining Fig. 1, we have used the constraints on fermion mass fit which fixes the VEV ratios  $|v_u/v_d| \simeq 70.3$ and $v_u'/v_d' \simeq |18.1 + 3.7 i|$~\cite{Babu:2020tnf}. 

{\bf Axion couplings and its hybrid nature:}
To study other phenomenological implications of the model as well to show its hybrid nature, we calculate the couplings of the axion to two photons, electron and the nucleon. For the axion-photon coupling we find 
\begin{eqnarray}
g_{a\gamma\gamma}=\frac{\alpha_{em}C_{em}}{2\pi f_a}~.
\end{eqnarray}
Here $C_{em}$=$\frac{8}{3}-1.92$ with the first term arising from triangle diagrams with fermions and the second model-independent term coming from the non-perturbative QCD effects such as $a-\pi^0$ mixing~\cite{Srednicki:1985xd,Georgi:1986df,GrillidiCortona:2015jxo,DiLuzio:2021pxd}. This coupling has the same value as in the DFSZ and KSVZ models. The axion coupling to electron and nucleon have the form given in Eq. (\ref{eq:Cf}) with the $C_{af}$-factors given by
\begin{eqnarray}
C_{ae} &=& \frac{24 r}{1+144 r} K_e \nonumber\\
C_{ap} &=& -0.47 + \frac{r}{1+144 r}(20.75 K_u -10.49 K_e) \nonumber \\
C_{an} &=& -0.02 + \frac{r}{1+144 r}(19.99 K_e - 9.73 K_u).
\label{coupling}
\end{eqnarray}
Here we have defined
\begin{equation}
K_u = \frac{2 {V_d}^2 + \tilde{v}^2}{v^2},~~~
K_e =  \frac{2{ V_u}^2 + \tilde{v}^2}{v^2}.
\label{kappa}
\end{equation}

\begin{figure}[t!]
		\centering
		\includegraphics[width=0.5\textwidth]{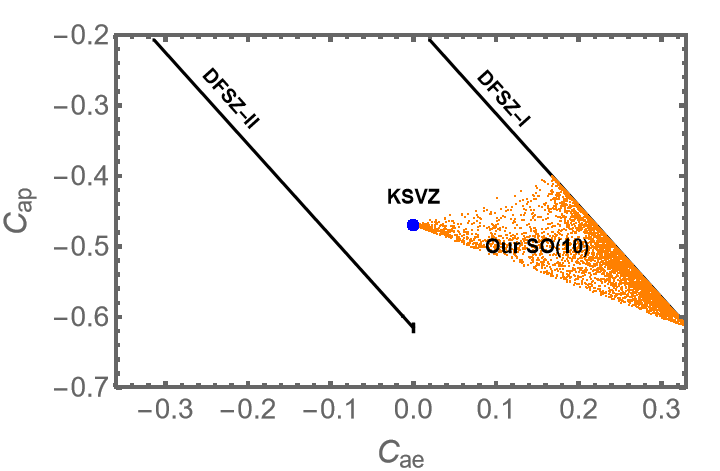}
  \includegraphics[width=0.5\textwidth]{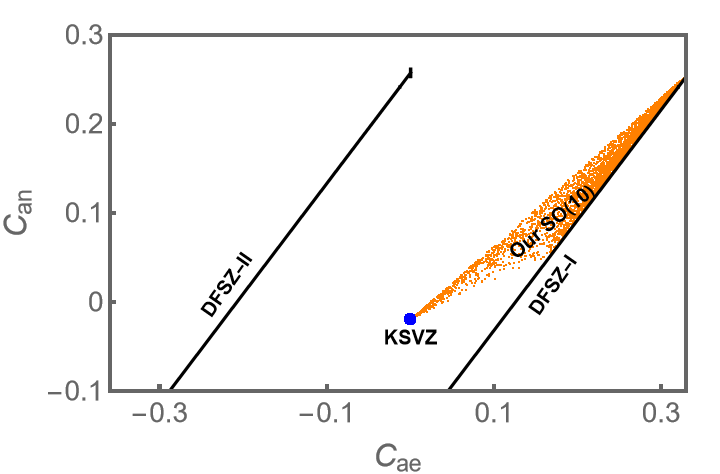}
	\caption{Correlations of axion-proton coupling (top) and axion-neutron coupling (bottom) with the axion-electron coupling in the $SO(10)$ model (yellow dots). Also shown are the values corresponding to KSVZ and DFSZ models. }
	\label{Fig:3}
\end{figure} 
 
We now turn to analyzing the model predictions and comparing them with the KSVZ and DFSZ axion models. In Fig. 2, we display the correlation between $C_{ae}$ and $C_{ap}$ (top figure) as well as $C_{ae}$ and $C_{an}$ (bottom figure)  as a way to test  the model. In obtaining Fig. 2, we have used the values of $|v_u/v_d|$ and $v_u'/v_d'$ obtained from fermion fits quoted above.  We find that
$K_e$ has a range $(1.5-2)$ corresponding to $\tilde{v} = V_u$ and $\tilde{v} \ll V_u$. The value of $K_u$ can be much smaller, with an upper limit of 0.5 (corresponding to $\tilde{v} = V_u$). This gives a range $C_{ae} = (0.25-0.33)$ for $r=(0.1-10)$. In the DFSZ-I model the model-dependent couplings of $(u,d,e)$ to the axion are given as $c_u^0= (1/3) \cos^2\beta$, $c_d^0 = c_e^0 =  (1/3) \sin^2 \beta$. The value of $c_e^0$ changes in DSFZ-II model to $c_e^0 = -(1/3)\cos^2\beta$~\cite{DiLuzio:2020wdo,GrillidiCortona:2015jxo}. 

The blue dots in Fig. 2 are the predictions of the KSVZ model. The black lines are the predictions of the DFSZ-I and DFSZ-II models and the yellow dots are the predictions of our $SO(10)$ model.  Depending on where the measured values of these observables lie, our model can be distinguished from the KSVZ and DFSZ models. These model predictions can be tested in various experiments which probe $C_{ae}$ and $C_{ap}/C_{an}$ individually as a function of $m_a$. For example, the solar axions are being probed using electron and nuclear interactions and detected by atomic and nuclear absorptions, respectively~\cite{XENON:2020rca, PandaX:2017ock,Lella:2023bfb, Bhusal:2020bvx}. Beam dump experiments are also able to probe these couplings separately using electron scattering and nuclear absorption~\cite{CCM:2021jmk, Capozzi:2023ffu, Waites:2022tov}.
A positive signal in axion-photon conversion experiments~\cite{Sikivie:1983ip,CAST:2017uph,ADMX:2019uok, IAXO:2020wwp, XENON:2022ltv} will determine $f_a$, in which case measurement of any one of the couplings $C_{ae},\, C_{ap} $ or $C_{an}$ in beam dump or solar axion searches would test our model.

We note that our model interpolates between the KSVZ and DFSZ-I models as we change the ratio $r=v^2_S/v^2_T$, as can be seen from the axion composition of the model in Eq.~(\ref{eq:afield}). As $r\to 0$, axion mostly consists of the field $\eta_S$, while in the $r\to \infty$ limit it mostly consists of $\eta_T$. From Eq. (\ref{eq:Yuk}) it is clear that in the $r \rightarrow 0$ limit the axion couples predominantly to the vector-like quarks in the {\bf 10}-fermion resembling the KSVZ model, while in the $r \rightarrow \infty$ limit it couples to the SM quarks in the {\bf 16}-fermions, in analogy to the DFSZ-I model. (The DFSZ-II model is not realized in this $SO(10)$ setup.) This interpolation is also reflected in Fig. 2 and Eqs. (\ref{coupling})-(\ref{kappa}). We see that, as $r\to 0$, the couplings to $e, p$ and $n$ reduce to the KSVZ values, whereas for $r\to \infty$ they go to DFSZ-I values. This confirms that $r$ is indeed an interpolating parameter and for general $r$, the predictions for the above observables lie in the yellow region of Fig. 2 representing the hybrid nature of our model.

{\bf Domain Wall in the \boldmath{$SO(10)$} model:} We now address the domain wall issue in our model, which is generically problematic in axion models~\cite{Sikivie:1982qv}.  To get the number of domain walls in the $SO(10) \times U(1)_a$ model, we employ the prescription given in Ref.~\cite{Ernst:2018bib}: 
\begin{equation}
N_{\rm DW} = {\rm min ~ integer} \left\{\frac{1}{f_a} \sum_i n_i\, c_i\, f_i,~n_i \in {\cal Z} \right\}~
\label{DW1}
\end{equation}
where $a \equiv \sum_i c_i \eta_i$ with the coefficients $c_i$ given in Eq. (\ref{eq:afield}) and the definition of $f_a$ given in Eq. (\ref{eq:fa}). In our $SO(10)$ model, since there are seven $\eta_i$'s there are seven integer $n_i$'s.  It is easy to see that $N_{DW} = 1$ in the model, obtained by setting $n_S = 1$ and other $n_i$'s to zero. This poses no problem with early universe cosmology~\cite{Sikivie:1982qv}.

{\bf Other cosmological issues:} First we note that, with $f_a \leq 4.2 \times 10^{11}$ GeV, the axion can serve as dark matter of the universe with the correct relic density~\cite{Borsanyi:2016ksw}.  This assumes that the PQ symmetry breaking occurs after inflation ends, in which case there is no contradiction with current limits on iso-curvature fluctuations.

The three singlet fermions $N_{1,2,3}$ of the model have sub-eV mass arising from Planck suppressed operators. As such, these fields can impact BBN.  However, their interactions with the SM particles are via the exchange of the $U(1)_a$ gauge boson, which has mass of order $v_a \sim 10^{12}$ GeV. As a result, the $N_i$'s go out of equilibrium at $T_*\simeq v_a(v_a/M_{\rm Pl})^{1/3}\simeq 10^9$ GeV. Therefore their contribution to energy density at the epoch of BBN is small with $\Delta N_{eff}\simeq 0.13$. This is in agreement with current CMB observations, but will be tested in planned next generation CMB experiments.

The singlet $\chi$ in our model has a mass of order TeV, and since its interactions with the SM particles are very weak, it can potentially overclose the universe. A simple resolution of this is to give it a large mass, of order the PQ breaking scale, by introducing an $SO(10)$ singlet fermion $N_0$ with zero $U(1)_a$ charge, which does not affect the anomaly cancellation or any other property of the model. This fermion will have  a Yukawa coupling of the type $\chi N_0 S^*$, which would give it a Dirac mass of order $10^{12}$ GeV. 
Now, a dimension-six four-fermion operator of the form $\chi N_1N_1N_1/M^2_{\rm Pl}$ is allowed in the Lagrangian. Since the $N_1$ is light, the decay $\chi \to 3N_1$ will proceed with a width given by $\Gamma_\chi\sim (m^5_\chi/M_{\rm Pl}^4)/192\pi^3 \sim 10^{-20}$ GeV. Since the $\chi$ fermion decays when $\Gamma_\chi$ equals the Hubble expansion rate, the decay temperature can be estimated to be of order GeV, which will leave the successes of BBN unaffected.

In conclusion, 
we have constructed a minimal and fully realistic $SO(10)$ model, a first of its kind, for a high quality axion that allows the maximum PQ scale of $4\times 10^{11}$ GeV with no domain wall problem. The axion can be the dark matter of the universe. The model has the unique property that it interpolates between the popular KSVZ and DFSZ models and predicts the nucleon and electron couplings of the axion to be different from other models, making the model testable.

{\bf Acknowledgements:} 
KB and RNM wish to thank the Mitchell Institute for Fundamental Physics \& Astronomy at Texas A\&M university for its warm hospitality during the Mitchell workshop in 2023 where this work was initiated.
The work of KSB is supported by the U.S. Department of Energy grant No. DE-SC0016013, and that of BD by DOE grant No. DE-SC0010813.

\bibliographystyle{style}
 \bibliography{BIB.bib}


\onecolumngrid

\appendix
\setcounter{secnumdepth}{2}
\section{Identifying the axion field}
\label{app:LALP}

We parametrize the neutral Higgs fields, freezing their radial modes, as
\begin{eqnarray}
&~&H_u^{10} = \frac{v_u e^{i \alpha_u}} {\sqrt{2}} e^{i \frac{\eta_u}{v_u}},~~~H_u^{\overline{126}} = \frac{v_u' e^{i \alpha_u'}} {\sqrt{2}} e^{i \frac{\eta_u'}{v_u'}},~~~H_d^{10} = \frac{v_d e^{i \alpha_d}} {\sqrt{2}} e^{i \frac{\eta_d}{v_d}},~~~H_d^{\overline{126}} = \frac{v_d' e^{i \alpha_d'}} {\sqrt{2}} e^{i \frac{\eta_d'}{v_d'}}, \nonumber \\
&~&\tilde{H}^{10} = \frac{\tilde{v} e^{i \tilde{\alpha}}} {\sqrt{2}} e^{i \frac{\tilde{\eta}}{\tilde{v}}},~~~\Delta_R^{\overline{126}} = \frac{v_R e^{i \alpha_R}} {\sqrt{2}} e^{i \frac{\eta_R}{v_R}},~~~S = \frac{v_S e^{i \alpha_S}} {\sqrt{2}} e^{i \frac{\eta_S}{v_S}},~~~T = \frac{v_T e^{i \alpha_T}} {\sqrt{2}} e^{i \frac{\eta_T}{v_T}}~.
\label{eq:fields}
\end{eqnarray}
Here the $H$-fields are neutral components of Higgs doubelts of the SM, with $\tilde{H}^{10}$ being the up-type Higgs doublet from the real {\bf 10}-plet $H'$, while the others are SM singlet fields.  All parameters $(v_j,\,\alpha_j)$ as well as the fields $\eta_j(x)$ are real in Eq. (\ref{eq:fields}).  
The axion field should be orthogonal to the three Goldstone bosons eaten up by the massive gauge fields $(Z_\mu,\,X_\mu,\,V^{(a)}_\mu)$, where $X_\mu$ corresponds to the gauge boson of $U(1)_X$ with $SO(10) \supset SU(5) \times U(1)_X$ and $V_\mu^{(a)}$ is the vector boson associated with the $U(1)_a$.  Furthermore, the axion is orthogonal to the four massive pseudoscalar fields. The Goldstone fields are identified as:
\begin{eqnarray}
    G_Z &=&N_Z(v_u \,\eta_u+v'_u\,\eta'_u-v_d \,\eta_d -v'_d \,\eta'_d + \tilde{v} \,\tilde{\eta})\nonumber\\
    G_X&=&N_X(10\, v_R\,\eta_R+2\,v_u\, \eta_u+2\,v'_u\, \eta'_u-2\,v_d\, \eta_d -2\,v'_d\,\eta'_d + 2\, \tilde{v}\, \tilde{\eta})\nonumber\\
    G_a&=&N_a(-2\,v_R\, \eta_R +v_T \,\eta_T+12\,v_S \,\eta_S-2\,v_u\, \eta_u-2\,v'_u\, \eta'_u-2\,v_d\, \eta_d-2\,v'_d\, \eta'_d)~.
    \label{eq:Gold}
\end{eqnarray}
The massive pseudoscalar fields are identified from the Higgs potential given in Eq. (\ref{eq:pot1}) as:
\begin{eqnarray}
{\cal A}_1&=&N_1\left[\frac{\eta_u}{v_u}-\frac{\eta'_u}{v'_u} \right] \nonumber \\
{\cal A}_2&=&N_2\left[\frac{\eta_d}{v_d}-\frac{\eta'_d}{v'_d} \right] \nonumber \\
{\cal A}_3&=&N_3\left[2\frac{\eta_T}{v_T} + \frac{\eta_u}{v_u} - \frac{\tilde{\eta}}{\tilde{v}}\right]\nonumber\\
{\cal A}_4&=&N_4\left[2\frac{\eta_T}{v_T} + \frac{\eta_d}{v_d} + \frac{\tilde{\eta}}{\tilde{v}}\right]~.
\label{eq:pseudo}
\end{eqnarray}
Here the $N_i$-factors are normalization coefficients.  The axion field is now readilty identified and is given in Eq. (\ref{eq:afield}).

\end{document}